\documentclass[
 reprint,
 amsmath,amssymb,
 superscriptaddress,
]{revtex4-1}

\setcitestyle{authoryear,aysep{,},open={},close={},semicolon}

\usepackage{color} 
\usepackage[usenames,dvipsnames,svgnames,table]{xcolor}
\usepackage{cancel} 
\usepackage{lineno} 
\usepackage[utf8]{inputenc} 
\usepackage{csquotes} 
\usepackage[caption = false]{subfig} 
\usepackage{hyperref} 
\hypersetup{colorlinks=true, linkcolor=blue, filecolor=magenta, urlcolor=blue, citecolor=gray} 
\usepackage{graphicx} 
\usepackage{dcolumn} 
\usepackage{bm} 
\usepackage{textcomp}
\usepackage{upgreek}

\begin{document}

\title{Frictional anisotropy of 3D-printed fault surfaces}

\author{Tom Vincent-Dospital}
\email{vincentdospitalt@unistra.fr}
\affiliation{Université de Strasbourg, ITES UMR 7063, Strasbourg F-67084, France}
\affiliation{SFF Porelab, The Njord Centre, Department of physics, University of Oslo, N-0316 Oslo, Norway}

\author{Alain Steyer}
\affiliation{Université de Strasbourg, ITES UMR 7063, Strasbourg F-67084, France}

\author{François Renard}
\affiliation{The Njord Centre, Department of Geosciences, University of Oslo, N-0316 Oslo, Norway}
\affiliation{Université Grenoble Alpes, Université Savoie Mont Blanc, CNRS, IRD, IFSTTAR, ISTerre, 38000 Grenoble, France}

\author{Renaud Toussaint}
\email{renaud.toussaint@unistra.fr}
\affiliation{Université de Strasbourg, ITES UMR 7063, Strasbourg F-67084, France}
\affiliation{SFF Porelab, The Njord Centre, Department of physics, University of Oslo, N-0316 Oslo, Norway}

\date{\today} 

\begin{abstract}
The surface morphology of faults controls the spatial anisotropy of their frictional properties, and hence their mechanical stability. Such anisotropy is only rarely studied in seismology models of fault slip, although it might be paramount to understand the seismic rupture in particular areas, notably where slip occurs in a direction different from that of the main striations of the fault. To quantify how the anisotropy of fault surfaces affects the friction coefficient during sliding, we sheared synthetic fault planes made of plaster of Paris. These fault planes were produced by 3D-printing real striated fault surfaces whose 3D roughness was measured in the field at spatial scales from millimeters to meters. Here, we show how the 3D-printing technology can help for the study of frictional slip. Results show that fault anisotropy controls the coefficient of static friction, with $\mu_{S//}$, the friction coefficient along the striations being three to four times smaller than $\mu_{S\perp}$, the friction coefficient along the orientation perpendicular to the striations. This is true both at the meter and the millimeter scales. The anisotropy in friction and the average coefficient of static friction are also shown to decrease with the normal stress applied to the faults, as a result of the increased surface wear under increased loading.
\end{abstract}

\maketitle


\section{Introduction}

Faults in the Earth's crust are complex systems along which earthquakes nucleate and propagate (e.g., \cite{Wibberley_complex}). Faults hold structures and heterogeneities at all scales (\cite{fault_imaging, fault_imaging2, rock_surf_bandwidth}). While they are often simplified to their simplest two-dimensional description (i.e., the fault plane), increasing complexity is now added to faults models (e.g., \cite{Rice_complex}). It is indeed considered that, to fully understand seismicity in various areas, it is paramount to account for some disorder in the faults frictional properties such as secondary faulting, off-fault damage or roughness of the fault plane (\cite{hetero_1, hetero_2, hetero_3, hetero_4, interlocked_rough_faults, fault_imaging_3, roughness1, roughness2}). For instance, the volume of damaged rocks, during the activation of a fault, depends on the initial contact roughness (\cite{wear_volume}) and, thus, a fault with a stronger roughness presents a different energy budget than a flat fault, as more energy is converted into surface area energy. In particular, roughness encourages the triggering of local events, but is believed to prevent the propagation of large-slip earthquakes (\cite{roughness2}). Additionally, large scale roughness tends to inhibit the propagation of any rupture faster than the shear wave velocity of surrounding rocks (\cite{bouchon_faulting_2010}).\\
Another degree of complexity is more rarely considered when modelling geological contacts and fault slip: the possible anisotropy in their frictional properties.
Morphological anisotropy is a known feature of faults, notably impacting the seismic waves velocity in their vicinity (\cite{seismic_anisotropy_3, seismic_anisotropy_1, seismic_anisotropy_2}) or the mobility of natural and injected fluids (\cite{perm_anisotropy}) in the subsurface. Frictional anisotropy, interestingly, is also regularly studied in other fields than seismology, for instance the tribology of rubber tires (\cite{rubber_1, rubber_2}), the strength of advanced adhesives (\cite{aniso_adhesive}), or the mitigation of water condensation (\cite{Condensation}). It is also considered to play a major role in nature (\cite{bio_friction_aniso}), for instance in the motion of numerous animals (\cite{aniso_adhesive, snake_friction_aniso, butterfly}) and the hydration of some plants (\cite{grass, lotus_bamboo}). In most cases, frictional anisotropy derives from the existence of preferential topographical orientations on, at least, one of the contact surfaces (\cite{friction_aniso_general, friction_anisotropy_2}). The length scale for such structural directivity can be as small as micrometer (\cite{friction_aniso_micro}) to nanometer (\cite{friction_aniso_atom, friction_aniso_atom_2}).
\begin{figure}
  \includegraphics[width=1\linewidth]{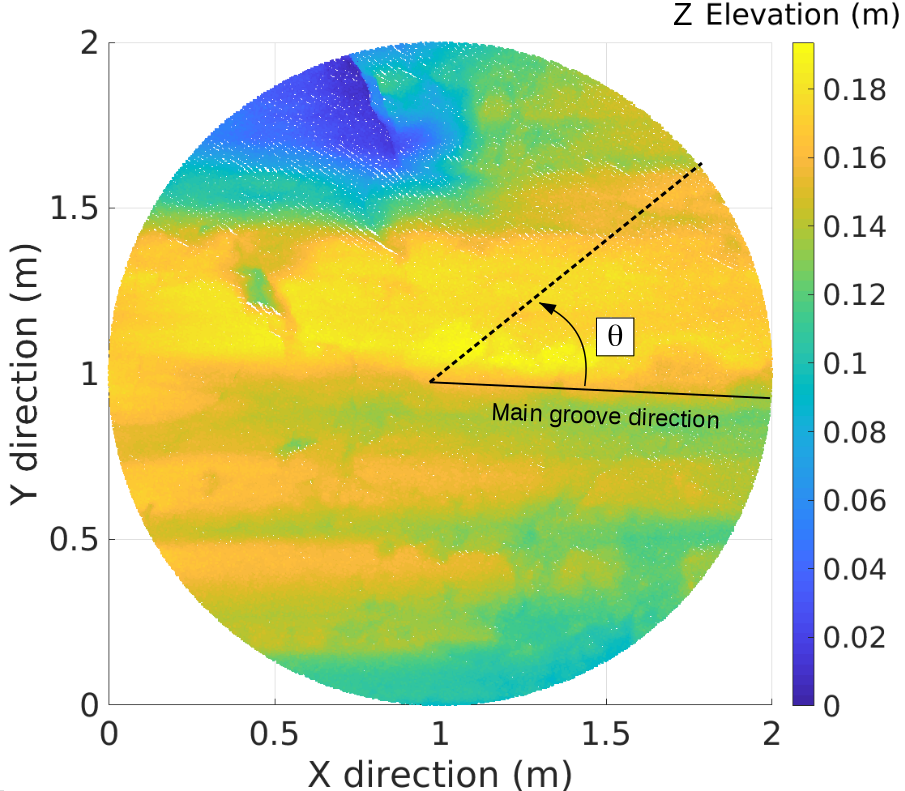}
  \caption{Topography (i.e. roughness) of the Corona Heights fault at the meter scale (\cite{fault_database, doi_fault_database}). This surface, called $S_{m}$, has a radius of $1$\,m and is defined on a $5$\,mm grid with a $1.25$\,mm elevation resolution. A parametric angle $\theta$ is defined from the main groove orientation.}
  \label{corona_metric}
\end{figure}
\begin{figure}
  \includegraphics[width=1\linewidth]{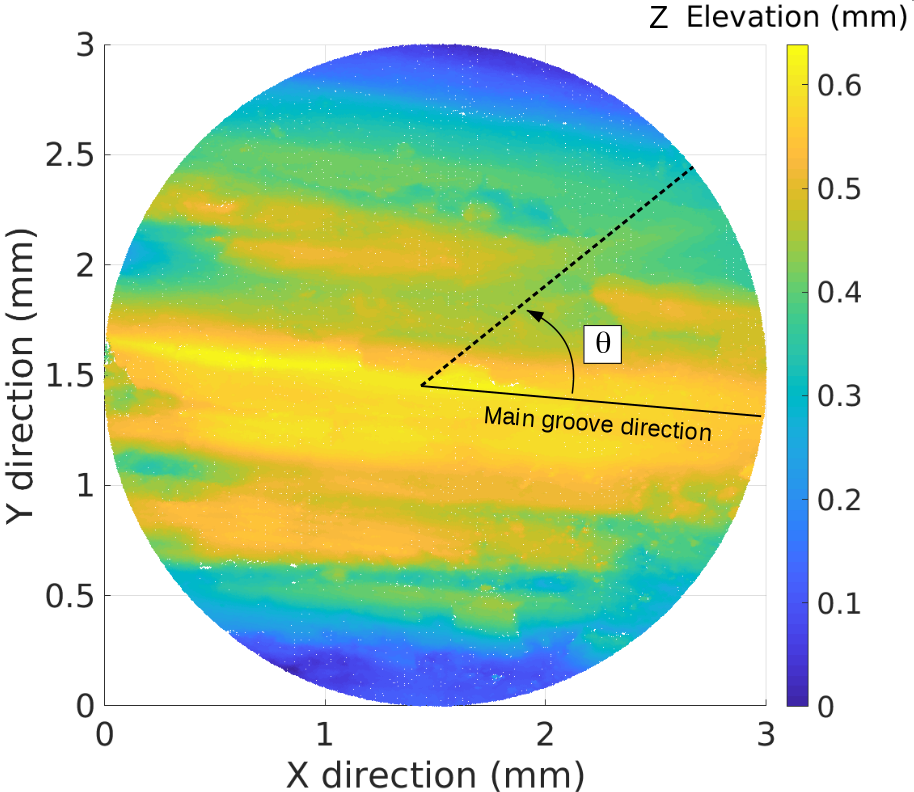}
  \caption{White light interferometry measurement of the topography of the Corona Heights fault plane at the millimeter scale (\cite{fault_database, doi_fault_database}). This surface, called $S_{mm}$, has a radius of $1.5$\,mm and is defined on a $2 \,\upmu$m grid with a $0.025\,\upmu$m elevation resolution. A parametric angle $\theta$ is defined from the main groove orientation.}
  \label{corona_millimetric}
\end{figure}
\begin{figure*}
  \includegraphics[width=0.8\linewidth]{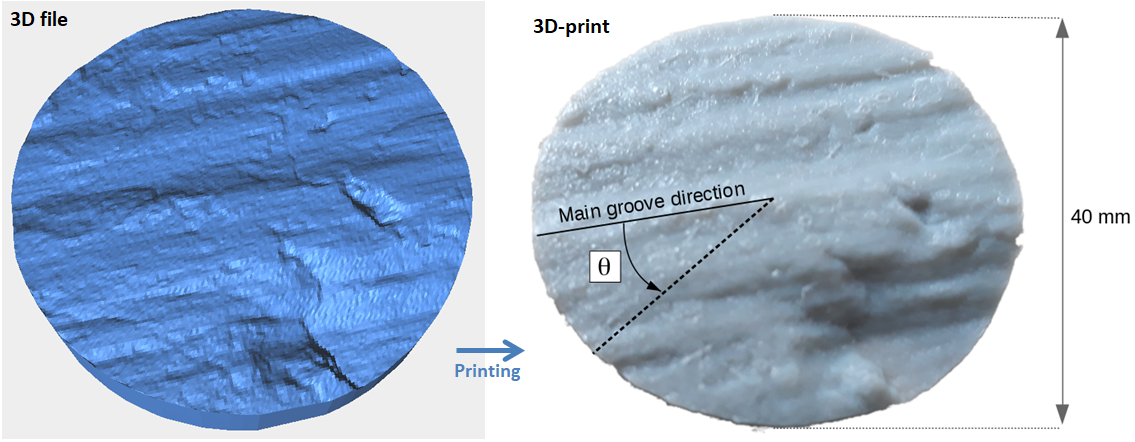}
  \caption{(Left): 3D file (STL file) obtained from the measured topography of the $S_{m}$ fault. The four 3D files generated for our friction experiments ($S_{m}$, $S_{mm}$ and their respective complementary) are available as Supplementary Material. (Right): Picture of a Polylactic acid 3D-print of the $S_{m}$ fault.}
  \label{PLAfault}
\end{figure*}
\\In seismic faults, such preferential orientations in their topography are observed at all scales (\cite{fault_imaging, fault_imaging2, rock_surf_bandwidth}) and originate from several processes. At the molecular level, rock forming crystals may display some frictional anisotropy. It is notably the case for antigorite, a mineral abundant in the Earth's upper mantle (\cite{friction_anisotropy_serp}). At the mesoscopic scale, the shear strength of foliated rocks is known to be anisotropic, due to the oriented planes in their constitutive mineralogy (\cite{strength_aniso1, strength_aniso2}). Fault zones in sedimentary basins are initiated by early fractures that often propagate in layered sediments. It can result in an anisotropic ramp-flat morphology of these fracture surfaces (\cite{rock_anisotropy}). For more mature faults having accumulated enough displacement, and above a given length scale (\cite{minimum_grooves}), the topography of the fault planes is also marked by slip induced wear, with striations and grooves of various wavelengths and amplitudes oriented along the main direction of slip (\cite{micro_grooves, macro_grooves}). If such morphological anisotropy of fault surfaces is well-known, its effect on the anisotropy of the frictional properties remains to be characterised.\\
Such a characterisation of frictional anisotropy could also be of interest for other types of rock contacts than strictly seismic faults, in particular for shallow rock joints and fractures, whose three-dimensional geometry is key in geotechnical engineering and for the structural stability of many man-made constructions (\cite{joint4, joint3, joint2, joint1}).\\
Here, we study how the morphology of faults controls the static coefficient of friction and the anisotropy of friction with regards to the main stress orientation during slip. To reach this goal, we produce 3D-prints of actual faults surfaces whose topography was measured in the field (\cite{fault_database}). We perform friction experiments with plaster of Paris casts of these 3D-printed faults. Results show that the coefficient of static friction along faults is highly anisotropic, a property that should henceforward be considered in numerical models of slip on seismic faults. We also show that this anisotropy is stress dependent, and should decrease with depth (e.g.,\,\cite{Byerlee1978}).

\section{3D printing and plaster casting of fault planes}

The actual morphology of natural faults can be difficult to assess, even if their long wavelength structures can be inferred by surface or subsurface imaging techniques (\cite{macro_grooves, imaging_sat, imaging_mag}). Yet, some fault planes are accessible to direct, high resolution, measurements, notably as they were exhumed by erosion and tectonic processes. For this study, we have used a series of digital fault surfaces. These fault roughness data were acquired with Light Detection And Ranging (LiDAR), laboratory laser profilometry, or white light interferometry techniques (\cite{fault_imaging2}). These data are available on an online public database (\cite{fault_database}), and in a repository with a doi number (\cite{doi_fault_database}). Should the reader hold some similar data, these authors welcome additions to this database. We have specifically selected fault roughness measurements performed on the Corona Heights fault (\cite{fault_imaging_3}) that outcrops near the Peixotto playground in San Francisco, California. These data cover surface areas with spatial scales in the range of millimeters to meters. Figures \ref{corona_metric} and \ref{corona_millimetric} show the fault surface at two spatial scales, one surface at the meter scale, defined on a $5\,$mm$\times5\,$mm grid, and one surface at the millimeter scale, defined on a $2\,\upmu$m$\times2\,\upmu$m grid. We will further on refer to these two surfaces as respectively $S_{m}$ and $S_{mm}$. Already, one can notice some preferential orientations in these topographies, and that the amplitude of fault roughness is, relatively to their size, somewhat larger at smaller scales\,($S_{mm}$) than at larger scales\,($S_{m}$) (\cite{fault_imaging_4}).
\begin{figure}[b]
  \includegraphics[width=1\linewidth]{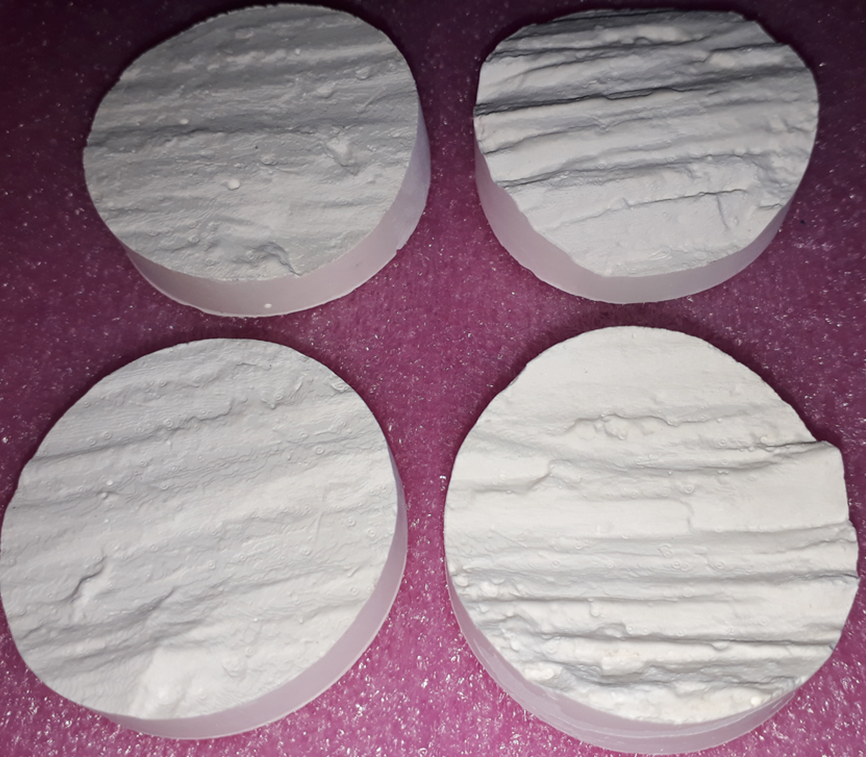}
  \caption{Plaster faults made using the 3D-printed moulds (i.e., an example of which is shown in Fig.\,\ref{PlasFaults}). Left: $S_{m}$ (top) and its complementary surface (bottom). Right: $S_{mm}$ (top) and its complementary surface (bottom). The samples have a diameter of $40$\,mm. One can appreciate that the fault at the millimeter scale ($S_{mm}$) shows a higher roughness aspect than at the meter scale ($S_{m}$).}
  \label{PlasFaults}
\end{figure}
\begin{figure*}
  \includegraphics[width=0.9\linewidth]{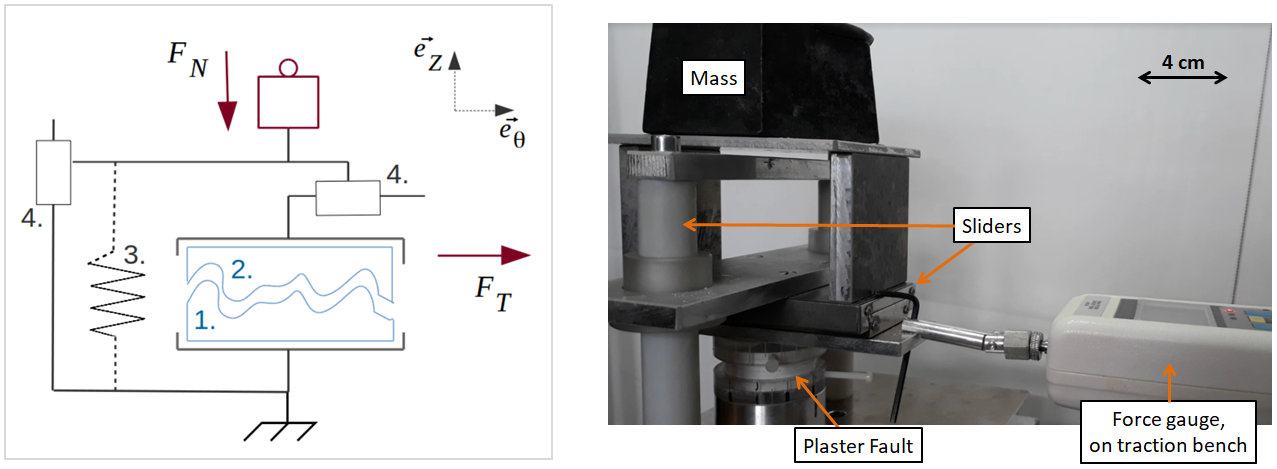}
  \caption{Schematic (left) and picture (right) of the shear apparatus. It contains: the complementary plaster-casted fault surfaces (1 and 2) installed between two clamps; the compression spring (3) necessary to obtain a null normal loading when required (not mounted on the picture); and two horizontal and vertical sliders (4) used to force the motion in the direction of interest while allowing for vertical displacement. The shear force $F_T$ is applied on the top fault wall in the $\boldsymbol{e_\theta}$ direction, while the bottom surface is kept fixed. It is measured by a Sauter\textsuperscript{\textregistered} FH500N force gauge (\cite{Sauter}). The normal force $F_N$ is applied by a dead weight that acts oppositely to the $\boldsymbol{e_Z}$ direction) on the top surface. For a visualisation purpose, the two plaster blocks (1 and 2) are offset vertically in the schematic, whereas in the experiments these two surfaces matched, with a quasi-null aperture.}
  \label{setup}
\end{figure*}
\\For our tests, we chose to limit these anisotropic surfaces to a circular sample geometry. We also applied a mild running-window median filter to smooth out spikes in the measured surfaces that could be associated to measurement noise. The window length of the filter was $10$ space steps, accounting for $5\,$cm for $S_{m}$ and for $20\,\upmu$m for $S_{mm}$. In order to run the friction experiments, we generated some opposing surfaces to the ones presented in Figs. \ref{corona_metric} and \ref{corona_millimetric}. These opposing surfaces could not be measured, as the actual fault walls that were facing $S_{m}$ and $S_{mm}$ are now eroded. To reconstruct them, we have applied the following transformation to the 3D coordinates $(X, Y, Z)$ of $S_{m}$ and $S_{mm}$:
\begin{equation}
\begin{split}
    &X' \sim X\\
    &Y' \sim -Y\\
    &Z' \sim -Z,
\end{split}
\end{equation}
where $X'$, $Y'$ and $Z'$ are the coordinates of the generated opposing surfaces and $(X, Y)$ give the map location of a given surface point of elevation $Z$, as represented in Fig.\,\ref{corona_metric}. We have thus assumed that the missing fault walls are complementary to the measured ones, so that, when pressed together before the friction tests, they form a bulk with negligible aperture between the two blocks. Such assumption for natural faults would only be partly verified. When having accumulated enough slip, a granular layer of gouge material may there have formed, and the two opposing sliding surfaces may not always perfectly match. However, our assumption is relevant for the youngest faults with a small amount of slip. We have also assumed that erosion did not significantly alter the fault plane, such that the measured topography is representative of the one of an actual buried fault. For the Corona Heights fault, this assumption is valid because the fault offsets silica-rich chert rocks with a high resistance to weathering.
\\After having obtained the surfaces, we isotropically (i.e., with the same factor in all directions) down- or up-scaled $S_{m}$ and $S_{mm}$ to fit a standard $4$\,cm diameter disk that matches the clamp size of our shear deformation apparatus. We also re-gridded the surfaces to match the lateral resolution of our 3D-printer (Ultimaker\textsuperscript{2} Extended+\,(\cite{Ultimaker})) that has a nozzle size of $250\,\upmu$m. The four surfaces (two fault surfaces and two opposing surfaces) were then 3D-printed into polylactic acid (PLA) material, as shown in Fig.\,\ref{PLAfault}. 
It should be noted that, even when designed to be flat, printed objects can present a natural roughness (\cite{3dprint_topo}), at a scale however smaller that the grooves observed on the printed faults. These intrinsic imperfections shall be comparable to $60\,\upmu$m, the elementary thickness of the PLA layers deposited by our 3D printer. In comparison, the 2D standard deviation of the elevation in our printed objects topography are $0.66\,$mm for $S_{m}$ and $1.7\,$mm for $S_{mm}$. The maximal elevation of these objects are, respectively, $3.7$\,mm and $8.3$\,mm. We thus consider that the small scale roughness ($\sim 60\,\upmu$m) from the printer's limit in resolution has a second order effect on the frictional properties of the surfaces.\\
Although we could have performed the friction experiments with the plastic pieces produced with the 3D printer, we have rather produced samples of plaster of Paris (gypsum) blocks moulded from the plastic faults. Plaster is known to be a reasonable model of porous brittle materials (\cite{plaster_modeling}), and the main goal of these casts was to work with a rock-like material, notably because plaster may wear and deform differently than plastic under shear. The fragile nature of plaster, and the potential friction-induced wear that the plaster was subjected to in our experiments, made us use new casts for each experimental realisation. The casts were generated with the following protocol: five volumes of water and eight volumes of powder of plaster of Paris were mixed and poured over the plastic moulds, then let to dry during one and half hour. The moulds, an example of which is shown in Fig.\,\ref{PLAfault}, were sprayed before each cast with a thin layer of silicon grease to avoid some of the fine plaster details to stick to the plastic during the mould release. The last step in the casts preparation was to dry them in an oven at a temperature of $40^\circ$C for one hour. As a result, we produced fault planes in blocks made of plaster of Paris, as shown in Fig.\,\ref{PlasFaults}.

\section{Experimental set-up and experimental conditions}

The shear apparatus used to perform the friction tests is shown in Fig.\,\ref{setup}. The two complementary surfaces are pressed together and mounted one on top of the other between the clamps of the shear apparatus. A normal force $F_N$ is applied on the top surface by using adjustable weights. In addition, a spring system of stiffness $625$\,N\,m\textsuperscript{-1} allows, if desired, to compensate for the machine empty weight of $13.7$\,N (i.e., the normal weight transmitted to the friction surfaces by the machine top clamp and structure and the top cast when no extra mass is used). A tangential driving shear force $F_T$ is then applied to the top fault wall in a given direction of the $(X, Y)$ plane. The amplitude of the force is measured by a Sauter\textsuperscript{\textregistered} force gauge (\cite{Sauter}).
The shear orientation is defined by the angle $\theta \in [0^\circ\,\,360^\circ]$ from the orientation of main grooves on the fault surface, as defined in Figs. \ref{corona_metric} and \ref{corona_millimetric}. A horizontal mechanical slider makes sure that the friction is evaluated in the direction of interest only, and a vertical slider allows upward or downward displacement of the top surface. While these sliders would ideally be perfectly lubricated, we have estimated their frictional resistance at the sliding velocity of our experiments, $F_c = 4\,$N$\pm0.5\,$N by performing a friction test with no fault installed in the machine (that is, with only air between the two clamps represented in Fig.\,\ref{setup}).
\begin{figure}[b]
  \includegraphics[width=1\linewidth]{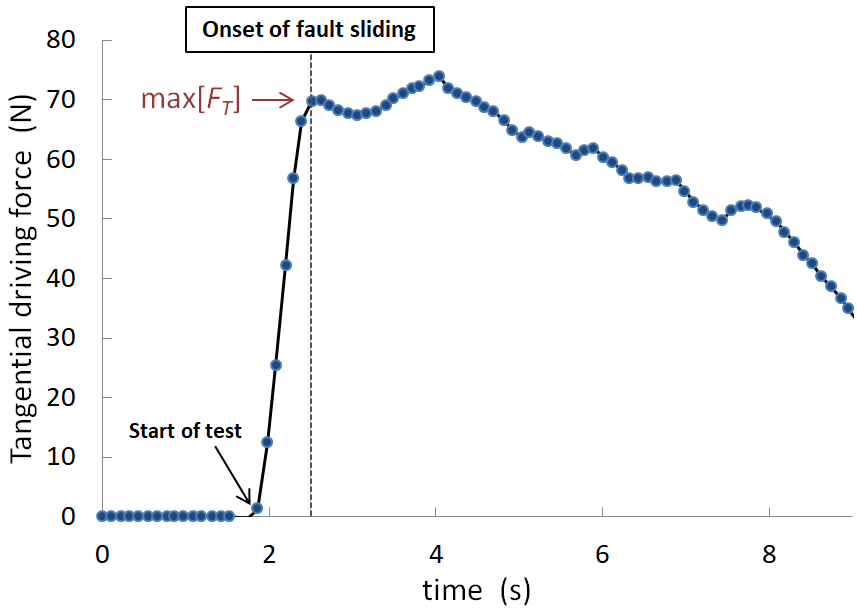}
  \caption{Typical tangential force versus time for a given slide angle ($\theta=30^\text{°}$) and a given normal force ($F_N=33.3$\,N) applied to the $S_m$ fault. From the local maximal value of the tangential force, at the onset of slip, a friction coefficient $\mu_s(33.3\,\text{N},\,30^\text{°})=2.0$ can here be calculated using Eq.\,(\ref{coulomb}).}
  \label{courbe}
\end{figure}
\\
The target speed of the test bench speed (that is, the demanded slip velocity) was fixed to a constant and equal to $1.3$\,mm\,s\textsuperscript{-1}. Of course, such a velocity may be orders of magnitude above that of typical tectonic solicitations (for instance, an ultra-fast oceanic ridge may reach an opening rate of $10$ to $20$\,cm\,yr\textsuperscript{-1} (\cite{Fast_ridge})). Here, we define the static friction in the experiments as the peak shear stress reached before sliding occurs, divided by the normal stress.\\
We characterise the anisotropy of this laboratory static friction coefficient for both the $S_{m}$ and $S_{mm}$ surfaces by performing a series of experiments where we vary the angle of loading with respect to the grooves. We ran friction tests every $30^\circ$ on both plaster faults. At each angle, the experiment was repeated at least three times in order to ensure that results are reproducible (with new casts each time, to avoid any wear related deviation). The standard deviation computed on these multiple measurements (typically $5$ to $10$\,N) was used to compute the error bars on our characteristics coefficients of static friction. $S_m$ was sheared under a normal stress $\sigma_N=10.9$\,kPa, while the tests performed on the rougher surface $S_{mm}$ were performed under $\sigma_N=26.5$\,kPa. A total of 76 experiments were performed, 37 using $S_{m}$ and 39 using $S_{mm}$.
At the onset of slip, the laboratory static friction coefficient is defined using a standard Coulomb's law (\cite{Bowden1951}):
\begin{equation}
    \mu_s(F_N, \theta) = \cfrac{\max\left[F_T(F_N, \theta)\right]-F_c}{F_N},
    \label{coulomb}
\end{equation}
where $\mu_s$ is the coefficient of static friction and $\max\left[F_T\right]$ is the (local) maximum tangential force applied at the onset of slip. In the following, we will also consider the mean driving and normal stresses, denoted $\sigma_T=F_T/(\pi r^2)$ and $\sigma_N=F_N/(\pi r^2)$, where $r=2\,$cm is the radius of the cast. Figure \ref{courbe} shows a typical measurement of a friction test, from which $\max\left[F_T\right]$, and hence $\mu_s$, are calculated.

\section{Results}

\subsection{Friction anisotropy}

\begin{figure}
  \includegraphics[width=1\linewidth]{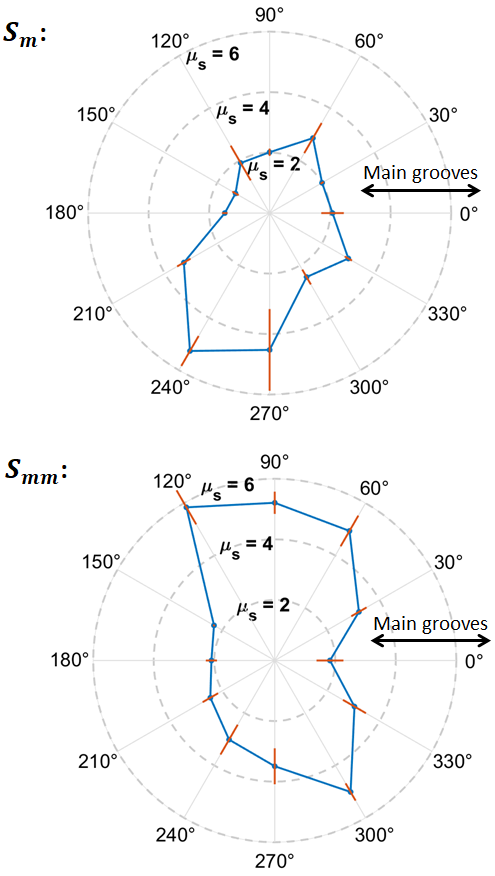}
  \caption{Coefficient of static friction of the $S_m$ fault (top) and $S_{mm}$ fault (bottom) as a function of the sliding direction $\theta$. Each transverse error bar is computed using at least three experimental realisations (with new plaster casts on each occurrence). Bar lengths display twice the standard deviation of the obtained coefficient. The tests performed with sample $S_m$ were done under a normal stress $\sigma_N=10.9$\,kPa, while those performed with the sample $S_{mm}$ (larger roughness) were done under $\sigma_N=26.5$\,kPa. The arrows indicate the orientation of the main grooves (see Figs.\,\ref{corona_metric} and.\,\ref{corona_millimetric}).}
  \label{results}
\end{figure}

The results are presented in Fig.\,\ref{results}. The derived coefficient of frictions are larger than $1$, which does not come as a surprise due to the non negligible roughness of our fault samples. Indeed, a large part of the contact area is bound to be perpendicular to the demanded slip (in any direction), inducing a strong resistance to motion.\\
On both fault samples, one can observe the strong anisotropy of the coefficient of static friction, with the maximum value of $\sigma_S$ being about four times larger than its minimum for $S_m$ and about three times larger for $S_{mm}$. In most experiments, the minimal friction is obtained along the main groove orientation (i.e., at $\theta=0^\circ$ or $\theta=180^\circ$), and the resistance to shear is larger perpendicularly to this orientation. The maximum of value is however never obtained exactly at $\theta=90^\circ$ or $\theta=270^\circ$ but rather along a neighbouring direction. Local maxima are indeed obtained for $\theta=60^\circ$ or $\theta=240^\circ$ when shearing $S_{m}$ and for $\theta=120^\circ$ or $\theta=300^\circ$ when shearing $S_{mm}$.

\subsection{Damage and stress dependence}

\begin{figure}
  \includegraphics[width=1\linewidth]{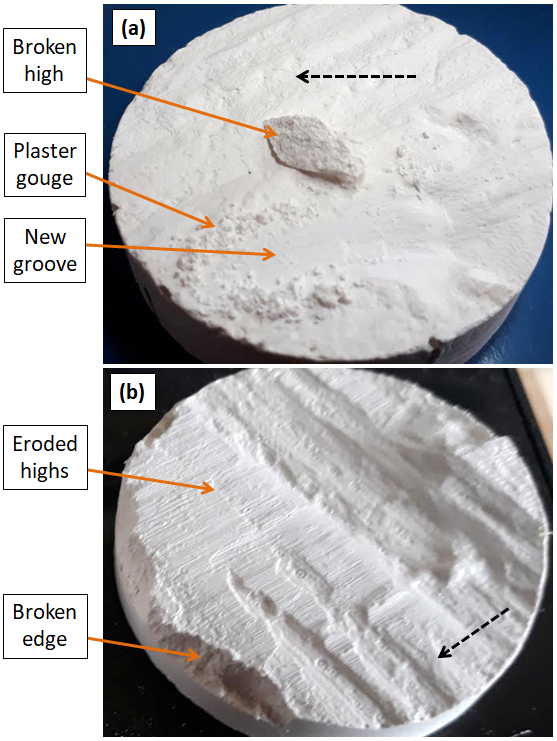}
  \caption{Various types of damage observed on the fault surfaces after the sliding tests. The dashed arrows show the slip directions of the complementary plaster casts during the tests. (a): $S_m$ sheared at $\theta=330^\circ$ and $\sigma_N=26.5$\,kPa. The indicated broken topographic high, on top of the fault, comes from the complementary surface. (b): Complementary of $S_m$ of a different experiment, sheared at $\theta=90^\circ$ and under a higher normal stress $\sigma_N=99.0$\,kPa. The visible damage is there less localised. Scale: the samples diameter is $40\,$mm.}
  \label{damages}
\end{figure}
\begin{figure}
  \includegraphics[width=1\linewidth]{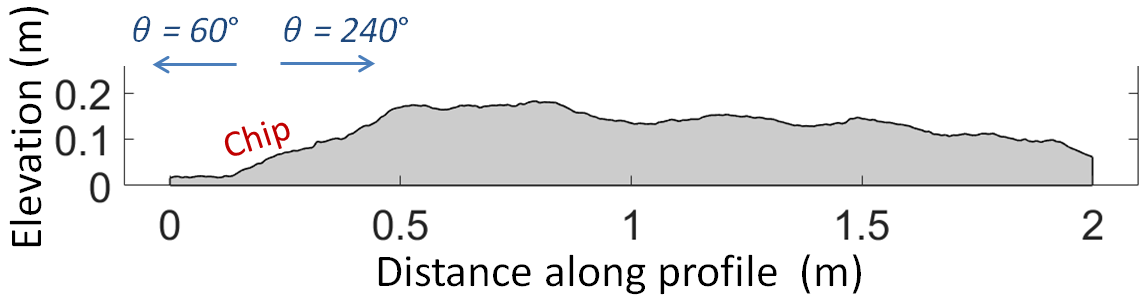}
  \caption{Elevation profile along the diameter of $S_m$ in the $60^\circ-240^\circ$ direction. The distances are reported in real word units and the axes are orthonormal. The arrows indicate the motion of the complementary surface for tests along $\theta=60^\circ$ and along $\theta=240^\circ$. The chip is only strongly interlocked in the latter orientation.}
  \label{profile}
\end{figure}
\begin{figure}
  \includegraphics[width=1\linewidth]{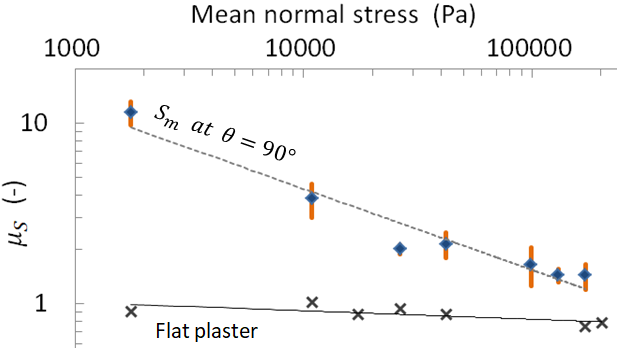}
  \caption{Coefficient of static friction $\mu_S$ as a function of the applied normal stress $\sigma_N$ for the fault $S_m$ sheared along the direction $\theta=90^\circ$ (squares). The length of the error bars is twice the standard deviation of $\mu_S$ obtained for three different tests. For reference, the straight dashed line indicates $\mu_S\propto {\sigma_S}^{-0.45}$. The static coefficient of friction measured on contacting flat plaster surfaces is shown (crosses) for comparison.}
  \label{stress_dep}
\end{figure}
\begin{figure}
  \includegraphics[width=1\linewidth]{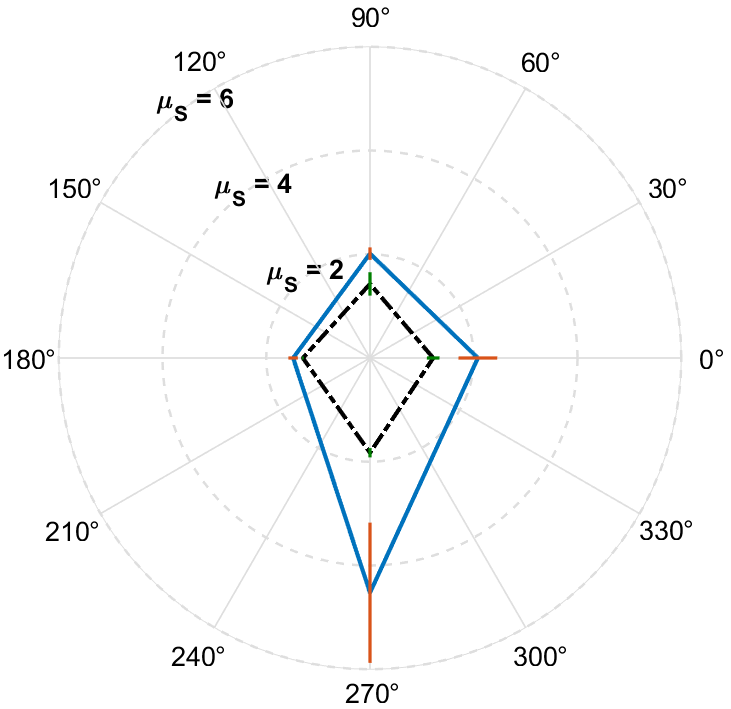}
  \caption{Coefficient of static friction $\mu_S$ of the $S_m$ fault along the cardinal directions $\theta=0^\circ,\,90^\circ,\,180^\circ \text{ and } 270^\circ$, for two normal stresses: $\sigma_N=10.9\,$kPa (outer plain line) and $\sigma_N=171\,$kPa (inner dashed line). The bars length is twice the standard deviation of the measured coefficient, for three different experimental realisations, at each angle and stress. Friction data at the two different normal loads do not overlap. One can notice the consistent reduction in the friction coefficient with a higher load and the reduced friction anisotropy at high $\sigma_N$.}
  \label{aniso_stress}
\end{figure}

Most of our experiments were destructive, with visible wear on the plaster samples after the shearing tests. This wear was the main reason calling for the production of new plaster casts for each experimental realisation, as we verified that repeating a same experiment with a previously used cast led to a significant (and here unwanted) drop in friction. The observed damage consists either in the formation of plaster powder (gouge) or in the rupture of topographic highs of the fault surfaces. Part of it might have initiated at the onset of the fault displacement (and hence be related to the static friction), while some of it has rather been induced by the subsequent sliding. Figure\,\ref{damages} shows some examples of these damage types.\\
Wear of seismic faults has been studied (e.g.,\,\cite{fault_wear, fault_damage_zones}) to, in particular, better understand the energy budget of the deformation, but also because this process may lubricate faults during slip (e.g.,\,\cite{fault_lubri,FluidPresMelt}) or modify the fault permeability to fluid flow (e.g.,\,\cite{gouge_flow1,gouge_flow2}). The present study focuses on the measurement of the coefficient of static friction and on its anisotropy, but we suggest that our 3D-print-based set-up could also enable the quantitative characterisation of damage during sliding along analogue fault surfaces.\\
We here keep to a qualitative assessment of which parts of the surfaces were mainly worn during each experiment. It seems that most of the shear resistance of the Corona Heights fault, at the millimeter scale ($S_{mm}$), arises from its grooves. By contrast, the friction of $S_{m}$ (representing a metric scale) is dominated by one chip in its field-scanned morphology (i.e., the main topographic low in Fig.\,\ref{corona_metric}). This chip being on the edge of the 3D-printed surface, but not on the edge of the real word fault, a finite size effect (inducing an artificial asymmetry of the fault wall) is certainly at play in the results reported in Fig.\,\ref{results} (top), notably explaining the strong asymmetry in $\mu_S$ for the opposite directions $\theta=60^\circ$ and $\theta=240^\circ$. Indeed, due to the finite size of our tested pieces, the interlocking of the chip is relatively free to unlock laterally along the $\theta=60^\circ$ direction, but is strongly locked by the surrounding plaster along the $\theta=240^\circ$ direction (see Fig.\,\ref{profile}).\\
Because the overall friction is likely to be affected by the surface wear, and because this wear is likely stress dependent, we have performed some friction tests on $S_m$ under various loads $\sigma_N$, at a given angle $\theta=90^\circ$. The results are shown in Fig.\,\ref{stress_dep}. At the highest tested stresses, $\mu_S$ seems, to an extent stress independent, with its mean variations lying within the measured error bars. While this result is compatible with the classical Coulomb theory (e.g.,\,\cite{Bowden1951}), one can observe, over a wider range of normal stresses, a consistent decrease in the friction coefficient with a higher normal stress. It could, in part, emanate from some limitations in our experimental set-up. For instance, at lower $\sigma_N$, the internal friction of the device ($F_c$) accounts for a more significant portion of the total measured tangential force, and the friction characterisation could thus be less accurate. We have however run similar tests on flat plaster surfaces, showing no significant variations in $\mu_S$ over the same stress range (see Fig.\,\ref{stress_dep}). The drop in friction coefficient with $\sigma_N$ is then likely related to the increased damage under higher normal loading (see Fig.\,\ref{damages}), reducing the overall shear resistance as asperities are easier to break. It might also result from the change in effective contact area with a higher load, with, proportionally, more pressure being borne by the surfaces parallel to the demanded motion, causing proportionally less of a resistance to slip. Such a drop of $\mu_S$ with the normal stress has already been reported for tilted contacts (\cite{Oded_Fineberg}), as the static coefficient of friction is not an absolute material constant (\cite{Oded_Fineberg, fractal_friction}).\\
We have then assessed the effect of the normal stress $\sigma_N$ on the anisotropy of the static friction coefficient. We performed frictional tests on the four poles of $S_m$ ($\theta=0^\circ,\,90^\circ,\,180^\circ \text{ and } 270^\circ$) at a higher load ($\sigma_N=171$\,kPa) than the load used before (i.e., $\sigma_N=10.9$\,kPa, as reported in Fig.\,\ref{results}). The newly measured coefficients of static friction are shown in Fig.\,\ref{aniso_stress}. One can notice the reduced friction anisotropy at high $\sigma_N$. The ratio between the maximum and the minimum value of $\mu_S$ indeed drops from $3.1$ at $\sigma_N=10.9$\,kPa to $1.5$ at $\sigma_N=171$\,kPa. This result suggests that the frictional anisotropy of faults is smaller at depth. This concept is naturally linked with the seminal \citet{Byerlee1978}'s law, stating that the roughness of fault planes (as well as the type of their constitutive rocks) has less effect on their maximum static frictional properties at larger depths.

\section{Discussion and conclusion}

Here, we show how the multiscale anisotropy of fault plane topography leads to an anisotropy in the frictional properties. Results confirm that seismic faults are prone to slide along some preferential orientations. The orientation that is the most likely is the one that faults have previously slid along, and which has shaped some guiding grooves in their morphology. Yet, displacements following other orientations are possible. Predicting the rupture direction of the next earthquake on a fault is thus not only dependent on assessing the main regional stress. The question should rather be along which orientation a rupture criterion (e.g., \cite{Bowden1951}) will first be exceeded. Such a subtlety might be of little importance for mature faults for which the stress principle orientations have not changed with time, because, in this case, the main stress is likely to act along the lowest coefficient of friction anyway. Yet, it could be paramount for faults under a changing geological load, where this alignment is not verified, or for immature faults, where the slip could be mainly governed by the anisotropy of early surfaces (i.e., where the slip does not coincide with the stress principal orientation, but is non-associated). Examples of slickensides (i.e., fault planes) commonly exist with several overlapping striations orientations (e.g.,\,\cite{multiple_orientations}), with rake and striations oblique to the actual orientation of the fault plane\,(e.g.,\,\cite{oblique_striations}). These observations indicate that the original slip direction (if assumed to initiate following Andersonian criterion (\cite{Anderson_faulting})) does not completely determine the direction of the next episode. Earthquakes occurring along abnormal directions (i.e., not in agreement with the local stress state) have been observed (e.g.,\,\cite{Balochistan_earthquake, tensile_Torishima_EarthQ, Anderson_faulting}), and their understanding might be eased by accounting for the possible frictional anisotropy of their surfaces (\cite{oblique_slip, stress_inversion_aniso}).\\
Note that frictional anisotropy should not only be considered at the full fault scale, but this property may spatially vary along the fault walls. Analytical solutions demonstrate that the stress around a fault is perturbed by its roughness, and a local slip can occur much before the entire fault is under yielding conditions in a given direction (\cite{interlocked_rough_faults}).\\
While we have here only measured the static coefficient of friction, we suggest that similar studies could be performed to characterise the coefficient of dynamic friction (i.e., by analysing the evolution of the resistance to motion, after the plaster faults start moving, as a function of the sliding orientation). Hence, not only the initial slip direction of an earthquake could be impacted by frictional anisotropy, but the complete slip trajectory (\cite{friction_anisotropy_2}). Changes in the slip direction within single earthquake rupture events are indeed sometimes observed, notably from bent grooves on post-mortem fault walls (\cite{variable_rake}).\\
We have, additionally, measured how the anisotropy in friction becomes less significant when the normal stress acting on a fault increases (i.e., with the fault depth), in general agreement with Byerlee's law (\cite{Byerlee1978}).
Such an effect likely derives from the stress related changes in rupture rheology and in damage type. The transition from a highly anisotropic to a relatively isotropic regime should typically occur when local stresses on the fault reach the yield strength of the material, $\sigma_y\sim5$\,MPa in the case of plaster (\cite{plaster_modeling}). This is about two orders of magnitude above the transition $\sigma_N\sim100\,$kPa at which we observed a strong reduction in anisotropy (see Fig.\,\ref{aniso_stress}), but our computed $\sigma_N$ is an average value which does not account for the potential strong stress concentrations at play in our faults. 
Considering that the strength $\sigma_y$ of rocks (e.g.,\,\cite{rock_strength}) is about two orders of magnitude (100 times) larger than that of plaster, fault frictional anisotropy could thus be only at play at pressures less than about $\sigma_N\sim100\times100\,\text{kPa}=10\,$MPa. This would correspond to the shallowest faults, at depths less than $\sigma_N/(g\rho)\sim500\,$m, where $\rho$ is the volumetric mass of rocks ($\sim2000\,$kg\,m\textsuperscript{-3}) and $g$ the gravity acceleration. Some care should however be taken when deriving such a conclusion by analysing resized samples as ours ($40\,$mm diameter samples representing meter or millimeter topographies), as the way matter breaks is length-scale dependent\,(e.g.,\,\cite{fault_imaging_4}).
Note also that other fault geometries than the one we have here studied may induce lesser stress concentration, so that a significant damage only occurs at a mean stress level directly comparable to the yield stress $\sigma_y$ of rocks. Thus, the possibility of frictional anisotropy should not be overlooked when studying fault buried up to $\sigma_y/(g\rho)\sim50\,$km. The heterogeneity of fault planes, and thus the anisotropy in this heterogeneity, may also still play a role under high stress, as roughness does not only encourage local yield, but also helps to suppress large slip events on moving faults (\cite{roughness2}).\\
Additionally to the assessment of the stability of (at least) shallow seismic faults, the characterisation of the frictional anisotropy of rock surfaces may be of importance in geotechnical engineering, for instance, for the stability of tunnels and foundations. There, the intrinsic strength anisotropy of foliated rocks is well studied (\cite{strength_aniso1, strength_aniso2}). Our work shows how one can also characterise the mechanical anisotropy of rough rock contacts, for instance, along joints (\cite{joint4, joint3, joint2, joint1}) and fractures (\cite{fracture1, fracture2}) between or inside rock formations.\\
A main point of this manuscript is, finally, to illustrate how the 3D-printing technology can help with new experimental designs in Earth Sciences, and this technology is getting a growing attention from the community (\cite{3dprint1, 3dprint2, 3dprint3, 3dprint_poro,3dprint_fault}), including the study of the frictional properties of 3D-printed fault analogues (\cite{3dprint_fault}). A direct continuation of the present work, for instance, could be to 3D-print and to test some faults surfaces beforehand filtered with various band-pass filters, in order to understand how the various wavelengths of the topography contribute to the global static friction coefficient, to the dynamical friction coefficient and to analyse the spatial distribution of the fault wear produced under various stresses and amounts of slip.\\

\subsection*{Acknowledgement}

\noindent
\textbf{Author contributions:} RT proposed the guidelines of this work, AS built the test machine and, with TVD, printed the faults and performed the friction experiments. FR advised on the mechanics of seismic faults. TVD wrote the first version of this manuscript and all authors contributed to the writing of its final version. We are grateful for the early experimental explorations performed by Marine-Sophie Jacob, Céline Fliedner, Aldo Mellado Aguilar, Gaëtan Leca and Laifa Rahmi, all students from the EOST/IPGS faculty at the University of Strasbourg. We also thank Amir Sagy from the Geological Survey of Israel for fruitful discussions, and acknowledge the support of the University of Strasbourg, of the IRP France-Norway D-FFRACT, and of SFF Porelab (project number 262644 of the Research Council of Norway). We thank the Strasbourg AV.Lab association for the use of their 3D printer.
\textbf{\\Author declarations:} The authors declare no competing interest. A funding support from the University of Strasbourg is acknowledged. Readers are welcome to comment and correspondence should be addressed to vincentdospitalt@unistra.fr or renaud.toussaint@unistra.fr.


\newpage
\bibliographystyle{plainnat_for_frontiers.bst}
\bibliography{scr_fault.bib}

\end{document}